\documentclass[twocolumn]{jpsj3}
\usepackage{txfonts}
\usepackage{color}

\title{Packing Fraction of a Two-dimensional Eden Model with Random-Sized Particles}

\author{Naoki Kobayashi\thanks{kobayashi.naoki75@nihon-u.ac.jp} and Hiroshi Yamazaki}
\inst{College of Industrial Technology, Nihon University, 2-11-1 Shin-ei, Narashino, Chiba 275-8576, Japan} %\\

\abst{
We have performed a numerical simulation of a two-dimensional Eden model with random-size particles.
In the present model, the particle radii are generated from a Gaussian distribution with mean $\mu$ and standard deviation $\sigma$.
First, we have examined the bulk packing fraction for the Eden cluster and investigated the effects of the standard deviation and the total number of
 particles $N_{\mathrm{T}}$.
We show that the bulk packing fraction depends on the number of particles and the standard deviation.
In particular, for the dependence on the standard deviation, we have determined the asymptotic value of the bulk packing fraction in the limit of the dimensionless
 standard deviation.
This value is larger than the packing fraction obtained in a previous study of the Eden model with uniform-size particles.
Secondly, we have investigated the packing fraction of the entire Eden cluster including the effect of the interface fluctuation.
We find that the entire packing fraction depends on the number of particles while it is independent of the standard deviation, in contrast to the bulk packing fraction.
In a similar way to the bulk packing fraction, we have obtained the asymptotic value of the entire packing fraction in the limit $N_{\mathrm{T}} \to \infty$.
The obtained value of the entire packing fraction is smaller than that of the bulk value.
This fact suggests that the interface fluctuation of the Eden cluster influences the packing fraction.
}

\begin{document}
\maketitle

\section{Introduction}

A rich phenomenon that particles of various sizes form clusters exists in nature.
Typical examples include wet granular materials,\cite{MN06} bacterial colonies,\cite{WIMM97} and food.\cite{SKM11}
One of the important factors in discussing the physical properties of these clusters is
 how much the inside of the cluster is filled with constituent particles.
Such a space-filling problem has a long history and has been studied both mathematically and from the point of view of physics.\cite{H06, BM60, B83, TS10}

In the past, numerical and theoretical studies have been made of cluster growth using discrete and continuous models.\cite{Meakin}
The Eden model was initially introduced by Eden in 1961 to describe the growth of biological cell colonies.\cite{E61}
It is a simple model that produces clusters that are compact but whose growing interface is comparatively rough because
 of the stochastic growth process.
In particular, its growing interface is known as a typical problem in the fluctuation statistics of a nonequilibrium system.\cite{Barabasi, Vicsek, T12}
Generally, these studies are applied to particles of equal size, although in the real world, the constituent particle sizes are distributed randomly.
However, hitherto there has been no theoretical study focusing on the situation where the particle size is random, although bulk properties of this model
 have been investigated quantitatively from the viewpoint of pattern formation such as the compact structure\cite{L85}
 and the packing fraction.\cite{WLB95}
Wang et al.\cite{WLB95} discussed the packing fraction of the Eden model with uniform-sized particles.
%%%%%%%%%%%%%%% Figure 1 %%%%%%%%%%%%%%%%%%%%%%%%%%%%%%%%%%%%%%%%%
\begin{figure}
\centering
\includegraphics[width=0.9\linewidth]{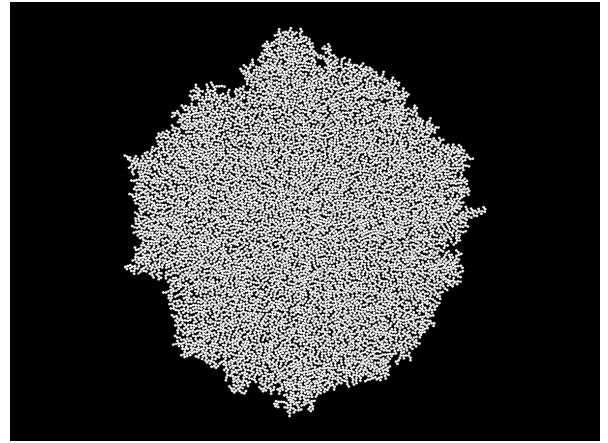}
\caption{Eden cluster with random-sized particles. The parameters are $N_{\mathrm{T}}=10000$, $\mu=1,$ and $\sigma=0.1$.}
\label{fig:Fig1}
\end{figure}
%%%%%%%%%%%%%%%%%%%%%%%%%%%%%%%%%%%%%%%%%%%%%%%%%%%%%%%%%%%%

In this paper, we report a numerical study of the packing fraction of a two-dimensional Eden model with random-sized particles that
 are chosen according to a Gaussian distribution.
The study is organized as follows.
In Sec. 2, we describe the growth rule of the model and the method of numerical simulation.
The result of numerical simulations is shown in Sec. 3.
Then, we also discuss the asymptotic values of the packing fraction based on our numerical data in Sec. 3.
A conclusion and future problems are given in Sec. 4.

\section{Growth Rule and Numerical Method}

The growth rule of the Eden model is very simple and the cluster growth is described as follows.
Let us start from a disk particle (seed) with a certain radius.
We prepare a new disk particle with a certain radius, place it in a direction chosen randomly with equal probability so as to be in contact with the seed,
 and it is incorporated into the cluster.
In the next, we randomly select one of those particles from the cluster with equal probability, and we place a new particle in a direction chosen
 randomly with equal probability so as to be in contact with the selected particle.
The above process is repeated without overlapping any particles in the cluster.
Note that if a particle overlaps with other particles in the clusters, these particles should be randomly selected again so that the Eden cluster grows without any particles overlapping.
The present growth rule belongs to Eden model version C.\cite{JB85}
However, while the original Eden model is lattice-dependent, the present model is an off-lattice one.

For random-sized particles characterized by a Gaussian distribution, the probability density function of particle radius $r$ is given by
\begin{align}
f(r;\mu, \sigma^2)=\frac{1}{\sqrt{2 \pi \sigma^2}}e^{-(r-\mu)^2 / 2\sigma^2},
\end{align}
where $-\infty < \mu < \infty$ and $\sigma > 0$ are the mean and standard deviation, respectively, of the radii.
In the present simulation, the particle radius is generated by applying the Box--Muller algorithm.\cite{BM58}
In addition, because the particle radius must be nonnegative, the parameters of the Gaussian distribution, 
especially the standard deviation $\sigma$, known as the {\it width parameter}, are adjusted so that the particle radius does not generate a negative value.
Figure 1 shows a typical two-dimensional example of the present model with the total number of particles $N_{\mathrm{T}}=10000$.
We examine the packing fraction of the present model by using our simulation data in the next section.

\section{Results and Discussion}
%%%%%%%%%%%%%%% Figure 2 %%%%%%%%%%%%%%%%%%%%%%%%%%%%%%%%%%%%%%%%%
\begin{figure}
\centering
\includegraphics[width=0.8\linewidth]{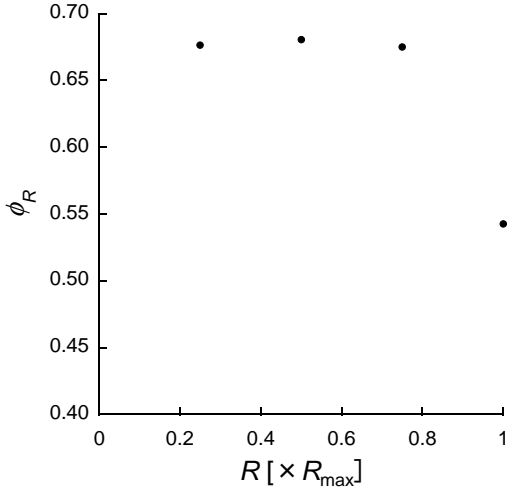}
\caption{Linear plot of $\phi$ vs $R$ for for random-size Eden model with $\mu = 20, \sigma = 2.0,$ and $N_{\mathrm{T}} = 50000$.}
\label{fig:Fig2}
\end{figure}
%%%%%%%%%%%%%%%%%%%%%%%%%%%%%%%%%%%%%%%%%%%%%%%%%%%%%%%%%%%%
%%%%%%%%%%%%%%% Figure 3 %%%%%%%%%%%%%%%%%%%%%%%%%%%%%%%%%%%%%%%%%
\begin{figure}
\centering
\includegraphics[width=0.8\linewidth]{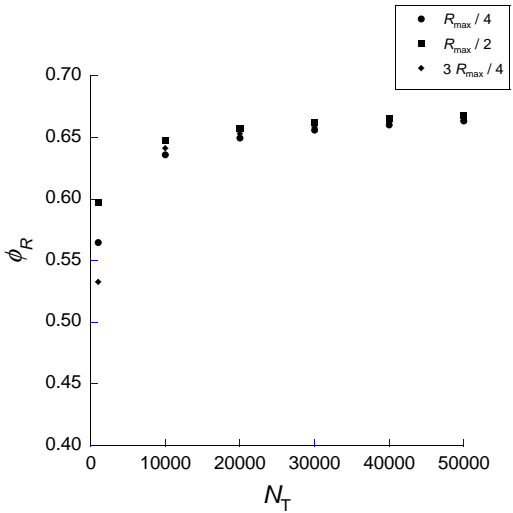}
\caption{Linear plot of $\phi$ vs $N_{\mathrm{T}}$ for random-size Eden model with $\mu = 20$ and $\sigma = 3.0$.
The data shown are  for $R = R_{\mathrm{max}} / 4,  R_{\mathrm{max}} / 2$, and $ 3 R_{\mathrm{max}} / 4$.}
\label{fig:Fig3}
\end{figure}
%%%%%%%%%%%%%%%%%%%%%%%%%%%%%%%%%%%%%%%%%%%%%%%%%%%%%%%%%%%%
%%%%%%%%%%%%%%% Figure 4 %%%%%%%%%%%%%%%%%%%%%%%%%%%%%%%%%%%%%%%%%
\begin{figure}
\centering
\includegraphics[width=0.8\linewidth]{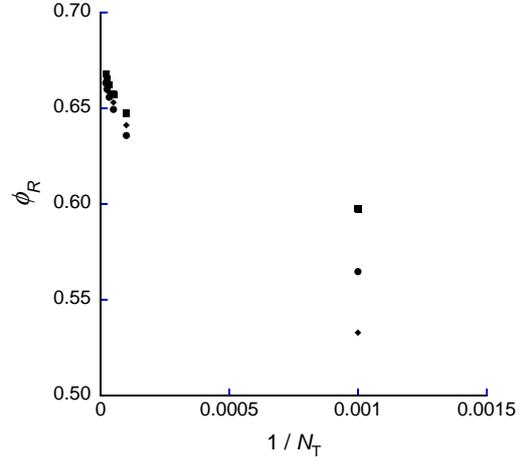}
\caption{$1 / N_{\mathrm{T}}$ plot of the estimated values of $\phi_R$ with $\mu = 20$ and $\sigma = 3.0$.
The symbols are the same as in Fig. 3.}
\label{fig:Fig4}
\end{figure}
%%%%%%%%%%%%%%%%%%%%%%%%%%%%%%%%%%%%%%%%%%%%%%%%%%%%%%%%%%%%
%%%%%%%%%%%%%%% Figure 5 %%%%%%%%%%%%%%%%%%%%%%%%%%%%%%%%%%%%%%%%%
\begin{figure}
\centering
\includegraphics[width=0.8\linewidth]{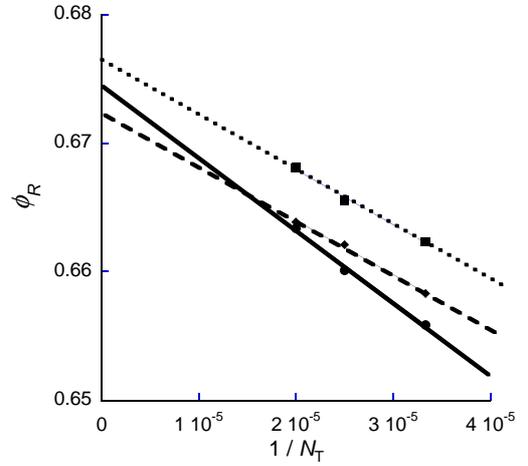}
\caption{Estimation of the linear fitting of the upper three plots in Fig. 4.
The symbols are the same as in Fig. 3.
The vertical intercepts are $0.674, 0.676$, and $0.672$ for $R = R_{\mathrm{max}} / 4$ (line),  $R_{\mathrm{max}} / 2$ (dotted line),
 and $ 3 R_{\mathrm{max}} / 4$ (dashed line), respectively.}
\label{fig:Fig5}
\end{figure}
%%%%%%%%%%%%%%%%%%%%%%%%%%%%%%%%%%%%%%%%%%%%%%%%%%%%%%%%%%%%

The packing fraction $\phi_R$ of the present model using the area of particles $A_{j} (j = 1, 2, \cdots, N_{\mathrm{T}})$ is defined as follows:
\begin{align}
\phi_R=\frac{\sum_{j} A_j}{A_{R}},
\end{align}
where $A_R$ is the area of the disk region with radius $R$ centered on the seed particle and the sum is taken over all particles existing in
 the disk region with radius $R$.
If there are particles crossing the boundary of the disk region, the packing fraction $\phi_R$ is computed including these areas.
In fact, whether or not particles crossing the boundary are included has little effect on the value of $\phi_R$.
Obviously, the closer $\phi_R$ is to 1, the denser the interior of the cluster.
In the following simulation results, the packing fraction $\phi_R$ averaged from 50 trials is shown.

Figure 2 shows linear plots of the packing fraction $\phi_R$ vs the radius $R$ with $\mu = 20, \sigma = 2.0,$ and $N_{\mathrm{T}} = 50000$.
Here $R_{\mathrm{max}}$ is the distance from the seed particle to the edge of the outermost particle.
The packing fractions $\phi_R$ calculated inside the Eden cluster are all larger than the packing fraction with $R = R_{\mathrm{max}}$.
In addition, there is little difference among the various radii $R$ inside the cluster.
This fact suggests that the fluctuation of the interface of the Eden cluster influences the packing fraction.
Therefore, for comparison with another model from the viewpoint of the space-filling problem, our discussion in Sect. 3.1 focuses on the packing
 fraction of the bulk structure of the Eden cluster.
The packing fraction including the effect of the interface fluctuation of the Eden cluster is discussed in Sect. 3.2.

\subsection{Bulk packing fraction of the Eden cluster}

Figure 3 shows the dependence of the packing fraction on the total number of particles with $\mu = 20$ and $\sigma = 3.0$.
It is confirmed that the packing fraction for each $R$ increases as the number of particles $N_{\mathrm{T}}$ increases.
However, it seems that the packing fraction saturates for large $N_{\mathrm{T}}$.
In fact, $\phi_{R=R_{\mathrm{max} / 4}} = 0.656, 0.660$, and $0.663$ for $N_{\mathrm{T}} = 30000, 40000$, and $50000$, respectively.
On the other hand, the curves for each $R$ in Fig. 3 are almost the same.
This result indicates that the packing fraction is almost independent of $R$ for large $N_{\mathrm{T}}$.
In Fig. 3, as the total number of particles increases, the packing fraction tends to asymptotically converge to a certain value.
In the field of critical phenomena, the $1/N_{\mathrm{T}}$ plot is known as a method for evaluating the asymptotic behavior of a certain value.\cite{G89}
Here, we evaluate the asymptotic value of the packing fraction $\phi_{R}$ using the $1 / N_{\mathrm{T}}$ plot.
Figure 4 shows the $1 / N_{\mathrm{T}}$ plot for the packing fraction with $\mu = 20$ and $\sigma = 3.0$ in Fig. 3.
We have made a linear fitting of the upper three plots, as shown in Fig. 5, and obtained an average value of $0.674$
 from the vertical intercepts of the three slopes in Fig. 5.
This asymptotic value of $0.674$ is almost the same as the value with large $N_{\mathrm{T}}$ in Fig. 3.
%%%%%%%%%%%%%%% Figure 6 %%%%%%%%%%%%%%%%%%%%%%%%%%%%%%%%%%%%%%%%%
\begin{figure}
\centering
\includegraphics[width=0.9\linewidth]{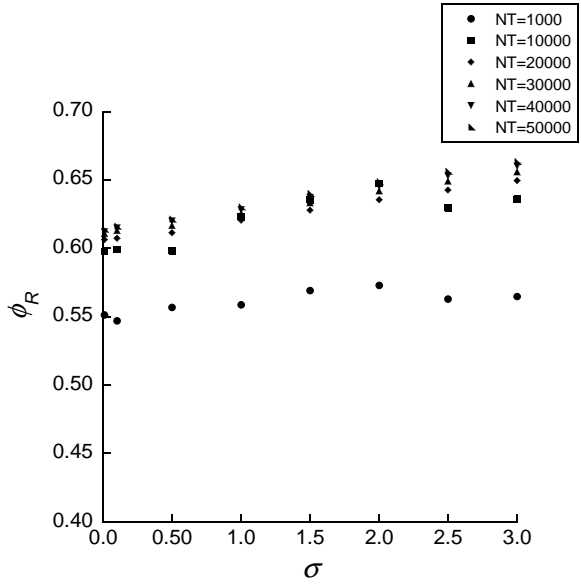}
\caption{Linear plot of $\phi_R$ vs. $\sigma$ for random-size Eden model with $\mu = 20$ and $R = R_{\mathrm{max}} / 4$. The data shown are for
 $N_{\mathrm{T}} = 1000, 10000, 20000, 30000, 40000$, and $50000$.}
\label{fig:Fig6}
\end{figure}
%%%%%%%%%%%%%%%%%%%%%%%%%%%%%%%%%%%%%%%%%%%%%%%%%%%%%%%%%%%%

%%%%%%%%%%%%%%% Figure 7 %%%%%%%%%%%%%%%%%%%%%%%%%%%%%%%%%%%%%%%%%
\begin{figure}
\centering
\includegraphics[width=0.9\linewidth]{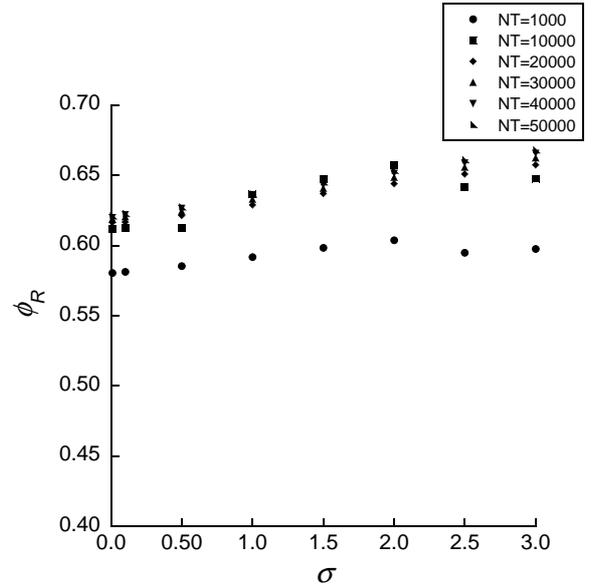}
\caption{Same as Fig. 6 for the radius $R = R_{\mathrm{max}} / 2$.}
\label{fig:Fig7}
\end{figure}
%%%%%%%%%%%%%%%%%%%%%%%%%%%%%%%%%%%%%%%%%%%%%%%%%%%%%%%%%%%%

%%%%%%%%%%%%%%% Figure 8 %%%%%%%%%%%%%%%%%%%%%%%%%%%%%%%%%%%%%%%%%
\begin{figure}
\centering
\includegraphics[width=0.9\linewidth]{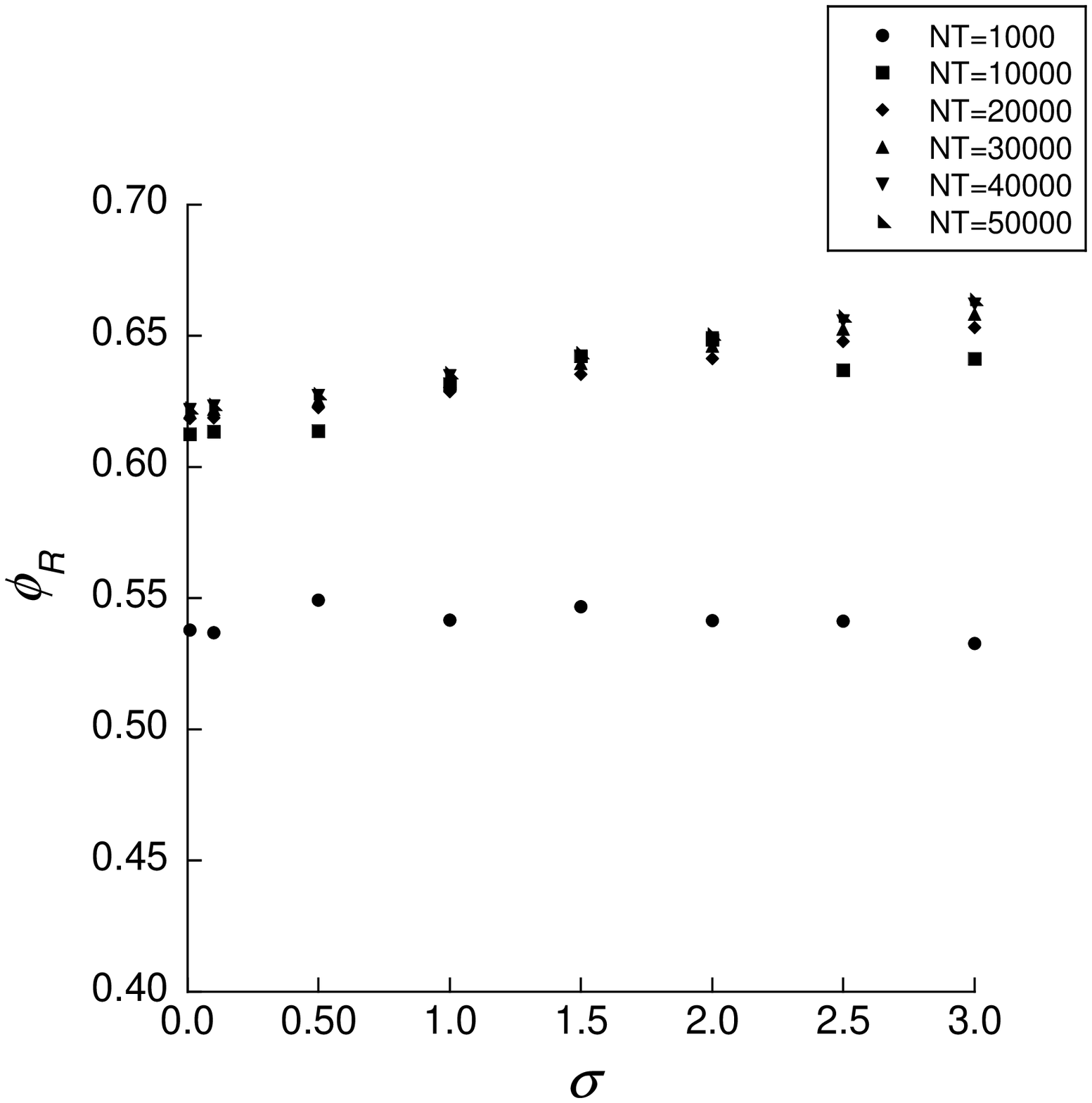}
\caption{Same as Fig. 6 for the radius $R = 3 R_{\mathrm{max}} / 4$.}
\label{fig:Fig8}
\end{figure}
%%%%%%%%%%%%%%%%%%%%%%%%%%%%%%%%%%%%%%%%%%%%%%%%%%%%%%%%%%%%

%Unlike the case of the standard deviation, 
Figures 6-8 show the dependence of the packing fraction on the standard deviation for $R = R_{\mathrm{max}}/4, R_{\mathrm{max}} / 2$, and $3 R_{\mathrm{max}} / 4$, respectively.
%Here $R_{\mathrm{max}}$ is the distance from the seed particle to the edge of an outermost particle.
These figures show the following similar trend.
%For $0 < \sigma < 3$, the values of the packing fraction $\phi_R$ with $N_{\mathrm{T}} = 20000, 30000, 40000$, and $50000$, have a positive correlation with $\sigma$.
As $\sigma$ increases, our data for large $N_{\mathrm{T}}$ except for $N_{\mathrm{T}}=1000$ and $10000$ show that $\phi_R$ has a positive correlation with $\sigma$.
For example, Fig. 9 shows the linear plot of $\phi_R$ vs $\sigma$ using the data for $N_{\mathrm{T}} = 30000$ in Figs. 6-8.
As $\sigma$ increases, the probability that a particle radius smaller than the average particle radius is generated increases.
Consequently, the packing fraction becomes large as these particles are placed inside the Eden cluster.
Because of the symmetric shape of the Gaussian distribution, particles larger than the average particle radius are also generated.
However, because these particles are difficult to place inside the cluster, they have little effect on the packing fraction.

%%%%%%%%%%%%%%% Figure 9 %%%%%%%%%%%%%%%%%%%%%%%%%%%%%%%%%%%%%%%%%
\begin{figure}
\centering
\includegraphics[width=0.9\linewidth]{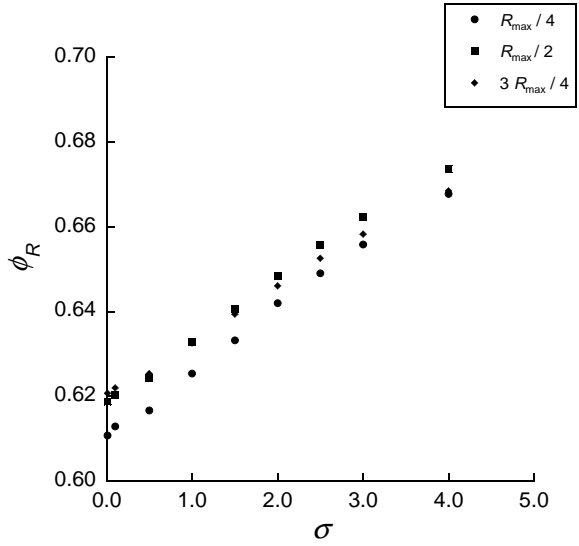}
\caption{Linear plot of $\phi_R$ vs $\sigma$ using the data for $N_{\mathrm{T}} = 30000$ in Figs. 6-8.}
\label{fig:Fig9}
\end{figure}
%%%%%%%%%%%%%%%%%%%%%%%%%%%%%%%%%%%%%%%%%%%%%%%%%%%%%%%%%%%%
%%%%%%%%%%%%%%% Figure 10 %%%%%%%%%%%%%%%%%%%%%%%%%%%%%%%%%%%%%%%%%
\begin{figure}
\centering
\includegraphics[width=0.9\linewidth]{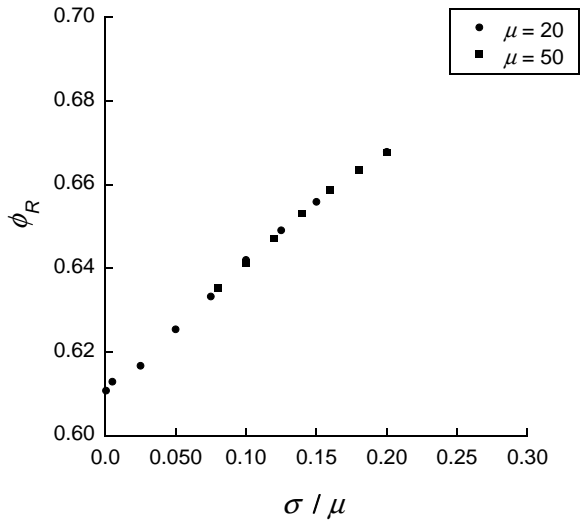}
\caption{Linear plot of $\phi_R$ vs. $\sigma / \mu$ with $N_{\mathrm{T}} = 30000$ and $R = R_{\mathrm{max}} / 4$.}
\label{fig:Fig10}
\end{figure}
%%%%%%%%%%%%%%%%%%%%%%%%%%%%%%%%%%%%%%%%%%%%%%%%%%%%%%%%%%%%
The value of $\phi_R$ satisfies $0 < \phi_R < 1$ on the basis of Eq. (2).
Thus, it is expected that $\phi_R$ gradually approaches an asymptotic value as $\sigma$ increases.
To obtain the asymptotic value of the packing fraction, it is necessary to simulate our model for large $\sigma$.
However, for large $\sigma$, a particle radius with nonphysical negative values may be generated by the Gaussian distribution.
If $\mu$ is increased, the particle radius does not become a negative value.
Because the packing fraction for the Eden cluster is independent of the average particle size $\mu$, we use the dimensionless
 standard deviation $\sigma / \mu$ instead of $\sigma$.\cite{SM68, OT81, YS88}
Figure 10 shows the linear plot of $\phi_R$ vs $\sigma / \mu$ with $N_{\mathrm{T}} = 30000$ and $R = R_{\mathrm{max}} / 4$ as an example.
From the figure, it appears that the packing fraction is gradually saturated as $\sigma / \mu$ increases.
However, note that a negative radius appears in the simulation when $\sigma / \mu > 0.20$.
Thus, in the same way as in Fig. 4, we have made a $1 / (\sigma / \mu)$ plot for the packing fraction
 and investigated the linear fitting of the upper three plots as shown in Fig. 11.
The vertical intercept of the line in Fig. 11 yields the asymptotic value of $\phi_R = 0.703$.
Similarly, we obtain $\phi_R = 0.710$ and $0.699$ from the vertical intercepts for $R = R_{\mathrm{max}} / 2$ and $3 R_{\mathrm{max}} / 4$, respectively.

%%%%%%%%%%%%%%% Figure 11 %%%%%%%%%%%%%%%%%%%%%%%%%%%%%%%%%%%%%%%%%
\begin{figure}
\centering
\includegraphics[width=0.9\linewidth]{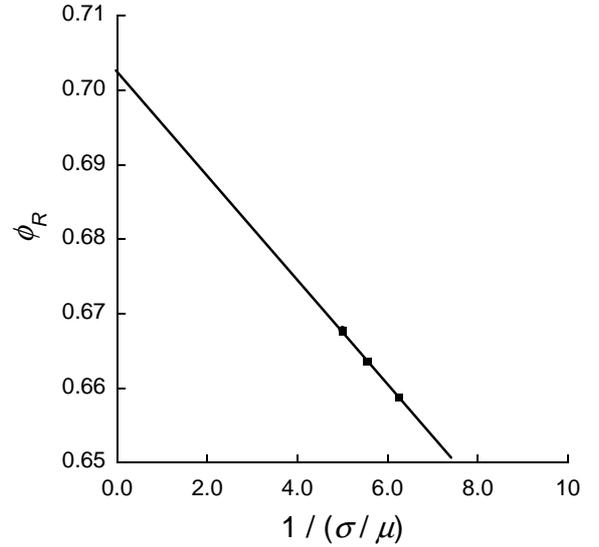}
\caption{Estimation of the linear fitting of the upper three plots of $1 / (\sigma / \mu)$ plot based on Fig. 10.}
\label{fig:Fig11}
\end{figure}
%%%%%%%%%%%%%%%%%%%%%%%%%%%%%%%%%%%%%%%%%%%%%%%%%%%%%%%%%%%%
%%%%%%%%%%%%%%% Figure 12 %%%%%%%%%%%%%%%%%%%%%%%%%%%%%%%%%%%%%%%%%
\begin{figure}
\centering
\includegraphics[width=0.9\linewidth]{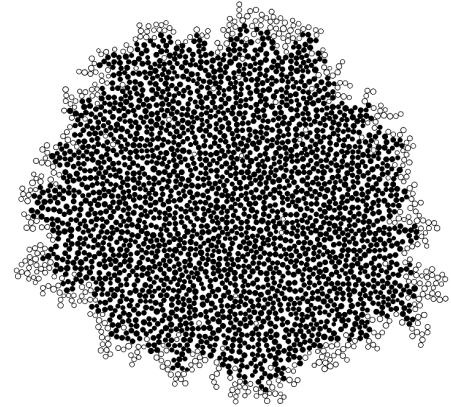}
\caption{Random-size Eden cluster with $N_{\mathrm{T}} = 3000$. The filled circles represent the particles constituting the Eden cluster with $N_{\mathrm{T}}=2500$.
On the other hand, the unfilled circles represent the particles generated between $N_{\mathrm{T}}=2500$ and $3000$.
Note that the circles exist in the bulk of the cluster, whereas the active particles exist on only the cluster boundary found by Wang et al.\cite{WLB95}}
\label{fig:Fig12}
\end{figure}
%%%%%%%%%%%%%%%%%%%%%%%%%%%%%%%%%%%%%%%%%%%%%%%%%%%%%%%%%%%%
These asymptotic values $\phi_{R}$ are considerably smaller than $\phi = 0.78$, which was conjectured for the two-dimensional random loose-packed limit.\cite{HFJ90}
On the other hand, the value of $\phi_{R}$'s are larger than $\phi = 0.547$ for our model evaluated in the random sequential adsorption (RSA) configuration,
 which is generated by sequentially placing disks of equal size randomly into a region without overlapping. \cite{HFJ86}
As a more direct comparison, $\phi = 0.65$ was obtained in a previous study of the two-dimensional Eden model by Wang et al.\cite{WLB95}.
While the particle radius in the study of Wang et al. was constant, that in the present model has the Gaussian distribution given by Eq. (1).
This fact implies that the particles in the present model tend to aggregate inside the Eden cluster as shown in Fig. 12.
From the viewpoint of the space-filling problem, Shi and Zhang simulated the random loose packing process of
 spherical particles whose radius has a Gaussian distribution.\cite{SZ08}
They reported that the standard deviation of the Gaussian distribution had little effect on the packing structure.
Our result is not consistent with the result of Shi and Zhang.
On the other hand, Sohn and Moreland experimentally measured the packing fraction of sand particles with radius having a Gaussian distribution.\cite{SM68}
They obtained the packing fraction as a function of the standard deviation.
In addition, the theoretical results obtained by Ouchiyama and Tanaka\cite{OT81} and Yu and Standish\cite{YS88} showed that the packing fraction of particles with radius having a Gaussian
 distribution also depends on the standard deviation.
The behavior of the packing fraction for our model is consistent with these results.

\subsection{Packing fraction and interface fluctuation}

In this subsection, we examine the effect of the interface fluctuation of the Eden cluster.
The packing fractions discussed above were based on Eq. (2).
Let us consider the packing fraction for the condition $R = R_{\mathrm{max}}$.

%%%%%%%%%%%%%%% Figure 13 %%%%%%%%%%%%%%%%%%%%%%%%%%%%%%%%%%%%%%%%%
\begin{figure}
\centering
\includegraphics[width=0.9\linewidth]{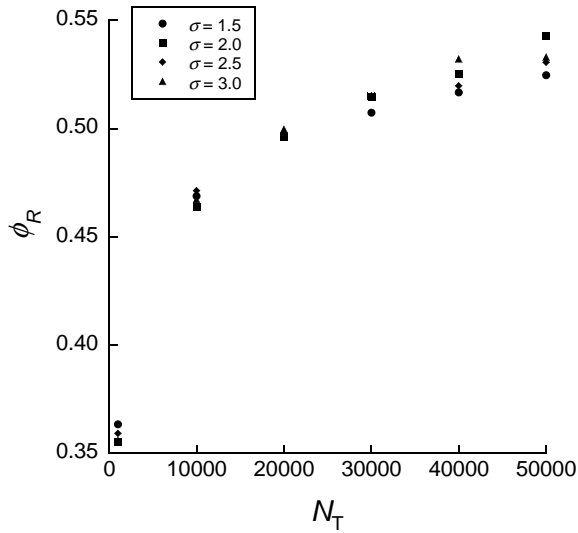}
\caption{Linear plot of $\phi_R$ vs. $N_{\mathrm{T}}$ for random-size Eden model with $\mu = 20$. The data shown are for various values of $\sigma = 1.5, 2.0, 2.5, 3.0$.}
\label{fig:Fig13}
\end{figure}
%%%%%%%%%%%%%%%%%%%%%%%%%%%%%%%%%%%%%%%%%%%%%%%%%%%%%%%%%%%%
%%%%%%%%%%%%%%% Figure 14 %%%%%%%%%%%%%%%%%%%%%%%%%%%%%%%%%%%%%%%%%
\begin{figure}
\centering
\includegraphics[width=0.9\linewidth]{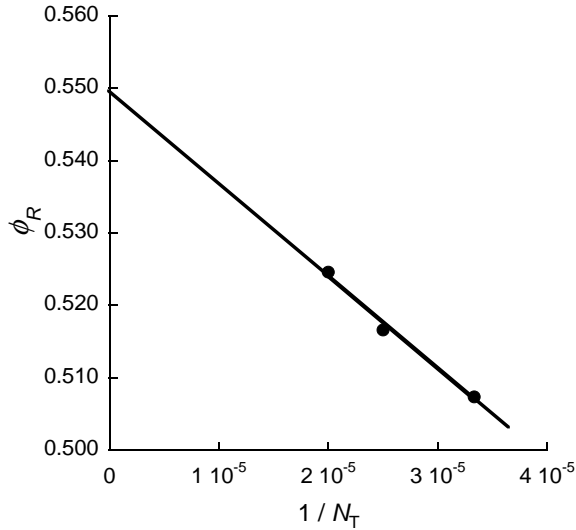}
\caption{Estimation of the linear fitting for the upper three $1 / N_{\mathrm{T}}$ plots with $\mu = 20$ and $\sigma = 1.5$ based on Fig. 13.}
\label{fig:Fig14}
\end{figure}
%%%%%%%%%%%%%%%%%%%%%%%%%%%%%%%%%%%%%%%%%%%%%%%%%%%%%%%%%%%%
Figure 13 shows the dependence of the packing fraction on the total number of particles for $\mu = 20$.
In a similar way to the bulk case, it is also confirmed that the packing fraction for each $\sigma$ increases as the number
 of particles $N_{\mathrm{T}}$ increases.
In addition, the curves for each $\sigma$ in Fig. 13 are all similar.
To obtain the asymptotic value of the packing fraction from the data in Fig. 13, we make a $1 / N_{\mathrm{T}}$ plot and investigate 
 the linear fitting of the upper three plots as shown in Fig. 14.
As a result, we obtain the value $\phi_{R} = 0.549$ from the vertical intercept of the line in Fig. 14.
In the same way, we obtain $\phi_R = 0.580$, $0.551$, and $0.563$ from the vertical intercepts for $\sigma = 2.0, 2.5$, and $3.0$, respectively.
This result implies that the packing fraction is almost independent of the standard deviation, unlike the bulk packing fraction.
In fact, Fig. 15 shows the dependence of the packing fraction on the standard deviation with $R = R_{\mathrm{max}}$.
For $0 < \sigma < 1$, the packing fraction $\phi_R$ has a slightly positive correlation with $\sigma$.
However, as $\sigma$ increases, our data for large $N_{\mathrm{T}}$ show that $\phi_R$ fluctuates around a constant value.
In the present model, most generated particles aggregate in the vicinity of the interface and contribute to the growth of the rough interface,
 which is a characteristic property of the Eden model (see Fig. 12).
Such interface fluctuation has a large influence on decreasing the packing fraction, as be seen from the definition in Eq. (2), and this influence is greater than
 the contribution to the packing fraction by $\sigma$.
Thus, the $\sigma$ dependence of the packing fraction is different from that inside the cluster.
%%%%%%%%%%%%%%% Figure 15 %%%%%%%%%%%%%%%%%%%%%%%%%%%%%%%%%%%%%%%%%
\begin{figure}
\centering
\includegraphics[width=0.9\linewidth]{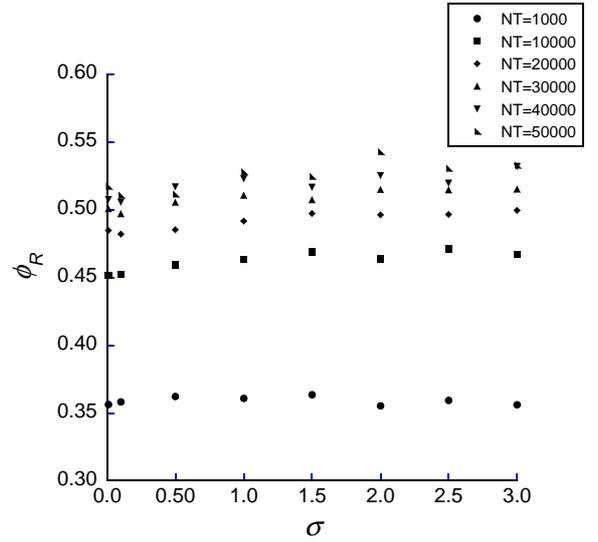}
\caption{Linear plot of $\phi_R$ vs. $\sigma$ for random-size Eden model with $\mu = 20$. The data shown are for
 $N_{\mathrm{T}} = 1000, 10000, 20000, 30000, 40000$, and $50000$.}
\label{fig:Fig15}
\end{figure}
%%%%%%%%%%%%%%%%%%%%%%%%%%%%%%%%%%%%%%%%%%%%%%%%%%%%%%%%%%%%

\section{Conclusion}

In this study, we have reported numerical results for the packing fraction $\phi_R$, defined by Eq. (2), of the Eden model with random-sized particles in two dimensions.
First, for comparison with another model from the viewpoint of the space-filling problem, we examined the bulk packing fraction for the Eden cluster.
It was confirmed that the bulk packing fraction depends on the number of particles and the standard deviation.
In particular, for the dependence on the standard deviation, we determined the asymptotic value of the bulk packing fraction to be $\phi_R = 0.703$ by evaluating the
 $1 / (\sigma / \mu)$ plot (see Fig. 11).
This value is larger than the packing fraction obtained in a previous study of the Eden model with uniform-sized particles.\cite{WLB95}.

Secondly, we discussed the packing fraction of the Eden cluster including the effect of the interface fluctuation.
The packing fraction with $R = R_{\mathrm{max}}$ depends on the number of particles.
In order to find the asymptotic value of the packing fraction with $R = R_{\mathrm{max}}$, we evaluated the $1 / N_{\mathrm{T}}$ plot (see Fig. 14).
Then, we obtained the asymptotic value $\phi_R = 0.549$, which is smaller than the bulk packing fraction.
This fact suggests that the interface fluctuation of the Eden cluster influences the packing fraction.
On the other hand, the packing fraction with $R = R_{\mathrm{max}}$ is independent of the standard deviation, in contrast to the bulk packing fraction.

In the present study, we assumed a Gaussian distribution as the fluctuation distribution for the particle size in the Eden cluster, but considering the phenomena
 of a complex system, it is natural to assume a lognormal distribution as the statistical property of the fluctuation distribution\cite{KKWM11} for bacterial
 cells,\cite{WKMM10} and food bolus,\cite{SKM11} and so form.
For a food bolus, it has been confirmed that the physical properties of the bolus and the size distribution of its constituent particles are closely related.
From this point of view, the Eden model discussed in the present study is a suitable model for theoretically and numerically analyzing it.
There have been previous studies on the packing fraction assuming a lognormal distribution in the random packing problem\cite{YMT96, HEC99} but not in the Eden model.
One of the most future tasks is to study physical quantities such as the packing fraction while changing the distribution function.

There is a possibility that the packing fraction of the present model can be further increased.
One way is to change the number of times $N_{\mathrm{B}}$ a direction is searched for when a new particle makes
 contact with a particle in the cluster.
In the present model, $N_{\mathrm{B}} = 1$.
We will show that the packing fraction increases as $N_{\mathrm{B}}$ increases in the future.

In Sect. 3, we suggested that the roughness of the Eden interface affects the packing fraction of the entire cluster.
Note that the packing fraction may change as a result of the poor convergence of the off-lattice Eden model.
Jullien and Botet \cite{JB87} simulated the off-lattice Eden model, where particles were selected with equal probability at any position where they are in contact with the growing
 cluster. 
That is, their off-lattice model was based on Eden model version A, which has poor convergence in the time evolution of its numerical simulation.\cite{JB85}
In this simulation, the active zone, which is the region where the new particles are aggregated, inside the cluster remains for a long time. \cite{Meakin}
On the other hand, in the model of Wang et al.\cite{WLB95} based on Eden model version C, the active zone inside the cluster has a relatively large probability
 of being eliminated.\cite{Meakin}
This indicates that the structure of the cluster in the model of Wang et al. tends to be denser than that in the model of Jullien and Botet.
In general, a growing Eden cluster satisfies the scaling relation $\xi_{\bot} \sim \overline{R}^{\beta}$, where $\xi_{\bot}$ is the width of the active zone \cite{PR84}
 and $\overline{R}$ is the mean radius of the cluster.
Note that the exponent $\beta$ is called the growth exponent \cite{FV85} if $\overline{R} \sim t$, where $t$ the time.
The model of Wang et al. yields $\beta \simeq 0.396$. \cite{WLB95, Meakin}
This value is consistent with the value of $\beta$ in Kardar--Parisi--Zhang (KPZ) universality. \cite{KPZ86}
On the other hand, the model of Jullien and Botet\cite{JB87} yields the value of $\beta \simeq 0.5$. \cite{Meakin}
Since $\xi_{\bot}$ describes the characteristics of the time evolution of the growing interface, one expects that $\xi_{\bot}$ is related to the packing fraction $\phi_R$
 of the cluster.
A future task will be to investigate the relation between the packing fraction and the other scaling exponents of the growing cluster.

\section*{Acknowledgment}

N.K. was supported in part by a Grant-in-Aid for Scientific Research (No. 15K00797) from the Japan Society for the Promotion of Science.


\begin{thebibliography}{99}
\bibitem{MN06}
N. Mitarai and F. Nori, 
Adv. Phys. {\bf 55}, 1 (2006).

\bibitem{WIMM97}
J. Wakita, H. Itoh, T. Matsuyama, and M. Matsushita,
J. Phys. Soc. Jpn. {\bf 66}, 67 (1997).

\bibitem{SKM11}
A. Sagawa, N. Kobayashi, and H. Moritaka,
Biosci. Biotechnol. Biochem. {\bf 75}, 2105 (2011).

\bibitem{H06}
T. C. Hales,
Discrete Comput. Geom. {\bf 36}, 5 (2006).

\bibitem{BM60}
J. D. Bernal and J. Mason,
Nature (London) {\bf 188}, 910 (1960) .

\bibitem{B83}
J. G. Berryman,
Phys. Rev. A {\bf 27}, 1053 (1983).

\bibitem{TS10}
S. Torquato and F. H. Stillinger,
Rev. Mod. Phys. {\bf 82}, 2633 (2010).

\bibitem{Meakin}
P. Meakin,
{\it Fractals, Scaling and Growth Far from Equilibrium} (Cambridge University Press, New York, 1998).

\bibitem{E61}
M. Eden, in {\it Proc. 4th Berkeley Symp. on Mathematical Statistics and Probability},
 ed. J. Neyman (University of California Press, Berkeley, 1961) p. 223.

\bibitem{Barabasi}
A.-L. Barab\'{a}si and H. E. Stanley, \textit{Fractal Concepts in Surface Growth} (Cambridge University Press, New York, 1995).

\bibitem{Vicsek}
T. Vicsek, \textit{Fractal Growth Phenomena} (World Scientific, Singapore, 1992) 2nd ed.

\bibitem{T12}
K. A. Takeuchi,
J. Stat. Mech., P05007 (2012).

\bibitem{L85}
F. Leyvraz,
J. Phys. A {\bf 18}, L941 (1985).

\bibitem{WLB95}
C. Y. Wang, P. L. Liu, and J. B. Bassingthwaighte,
J. Phys. A: Math. Gen. {\bf 28}, 2141 (1995).

\bibitem{JB85}
R. Jullien and R. Botet,
J. Phys. A: Math. Gen. {\bf 18}, 2279 (1985).

\bibitem{BM58}
G. E. P. Box and M. E. Muller,
Ann. Math. Stat. {\bf 29}, 610 (1958).

\bibitem{G89}
A. J. Guttmann,
in {\it Phase Transitions and Critical Phenomena}, ed. C. Domb and J. L. Lebowitz (Academic Press, London, 1989), Vol. 13, p. 1.

\bibitem{SM68}
H. Y. Sohn and C. Moreland,
Can. J. Chem. Eng. {\bf 46}, 162 (1968).

\bibitem{OT81}
N. Ouchiyama and T. Tanaka,
Ind. Eng. Chem. Fundam. {\bf 20}, 66 (1981).

\bibitem{YS88}
A. B. Yu and N. Standish,
Powder Technol. {\bf 55}, 171 (1988).

\bibitem{HFJ90}
E. L. Hinrichsen, J. Feder, and T. J\o ssang,
Phys. Rev. A {\bf 41}, 4199 (1990).

\bibitem{HFJ86}
E. L. Hinrichsen, J. Feder, and T. J\o ssang,
J. Stat. Phys. {\bf 44}, 793 (1986).

\bibitem{SZ08}
Y. Shi and Y. Zhang,
Appl. Phys. A {\bf 92}, 621 (2008).

\bibitem{KKWM11}
N. Kobayashi, H. Kuninaka, J. Wakita, and M. Matsushita,
J. Phys. Soc. Jpn. {\bf 80}, 072001 (2011).

\bibitem{WKMM10}
J. Wakita, H. Kuninaka, T. Matsuyama, and M. Matsushita,
J. Phys. Soc. Jpn. {\bf 79}, 095002 (2010).

\bibitem{YMT96}
A. Yang, C. T. Miller, and L. D. Turcoliver,
Phys. Rev. E {\bf 53}, 1516 (1996).

\bibitem{HEC99}
D. He, N. N. Ekere, and L. Cai,
Phys. Rev. E {\bf 60}, 7098 (1999).

\bibitem{JB87}
R. Jullien and R. Botet,
{\it Aggregation and Fractal Aggregates} (World Scientific, Singapore, 1987).

\bibitem{PR84}
M. Plischke and Z. R\'{a}cz,
Phys. Rev. Lett. {\bf 53}, 415 (1984).

\bibitem{FV85}
F. Family and T. Vicsek,
J. Phys. A: Math. Gen. \textbf{18}, 75 (1985).

\bibitem{KPZ86}
M. Kardar, G. Parisi, and Y.-C. Zhang,
Phys. Rev. Lett. \textbf{56}, 889 (1986).

\end{thebibliography}
\end{document}